\documentclass[conference]{IEEEtran}
\IEEEoverridecommandlockouts
\usepackage{cite}
\usepackage{amsmath,amssymb,amsfonts}
\usepackage{MnSymbol}
\usepackage{algorithmic}
\usepackage{graphicx}
\usepackage{textcomp}
\usepackage[table,dvipsnames]{xcolor}
\usepackage{microtype}
\usepackage{enumitem}
\usepackage{changepage}
\usepackage{booktabs} 
\usepackage{multirow}
\usepackage{listings}
\usepackage{xcolor}
\usepackage[most]{tcolorbox}
\usepackage{xcolor}
\usepackage[colorlinks=false]{hyperref} 
\usepackage[nameinlink]{cleveref}
\usepackage{tikz}
\usetikzlibrary{shapes.multipart, positioning, arrows.meta}
\usepackage{adjustbox}
\def\BibTeX{{\rm B\kern-.05em{\sc i\kern-.025em b}\kern-.08em
    T\kern-.1667em\lower.7ex\hbox{E}\kern-.125emX}}
\hypersetup{
pdfborder={0 0 0}
}
\usepackage{csquotes}
\usepackage{microtype}
\usepackage{svg}
\usepackage{makecell}
\usepackage{siunitx}
\usepackage{tabularx}
\usepackage{booktabs}
\usepackage{listings}
\usepackage{xcolor}

\definecolor{lstgray}{gray}{0.97}
\definecolor{lstframe}{gray}{0.55}

\lstdefinestyle{aptcjson}{
  language=,
  basicstyle=\ttfamily\footnotesize,
  columns=fullflexible,
  breaklines=true,
  breakatwhitespace=true,
  keepspaces=true,
  showstringspaces=false,
  frame=single,
  framerule=0.4pt,
  rulecolor=\color{lstframe},
  backgroundcolor=\color{lstgray},
  xleftmargin=0.6em,
  xrightmargin=0.6em,
  framexleftmargin=0.4em,
  framexrightmargin=0.4em,
  aboveskip=0.6em,
  belowskip=0.2em,
  tabsize=2
}

\definecolor{backcolor}{RGB}{226, 223, 223}
\usepackage{listings}
\lstdefinelanguage{JSON}{
  basicstyle=\ttfamily\tiny,
  keywordstyle=[1]\color{Mahogany},
    sensitive=true,
    stringstyle=\color{red},
    numbers=none,
    stepnumber=1,
    showstringspaces=false,
    tabsize=1,
    breaklines=true,
    breakatwhitespace=false,
    basicstyle=\scriptsize\ttfamily,
    print, 
    backgroundcolor=\color{backcolor},
    escapechar=@,
morekeywords={prioritizedPentestcases,assessedThreat,usedAttackVector,entryPoint,executedOn,resourceContainer,assemblyContext,asset,violatedSecurityProperties,id,name,href}
}

\crefname{lstlisting}{listing}{listings}   
\Crefname{lstlisting}{Listing}{Listings}   

\begin{document}



\title{Towards Leveraging LLMs to Generate Abstract Penetration Test Cases from Software Architecture}
\author{
\IEEEauthorblockN{
Mahdi Jafari\IEEEauthorrefmark{1},
Rahul Sharma\IEEEauthorrefmark{1},
Sami Naim\IEEEauthorrefmark{1},
Christopher Gerking\IEEEauthorrefmark{1},
Ralf Reussner\IEEEauthorrefmark{1}
}
\IEEEauthorblockA{\IEEEauthorrefmark{1}Karlsruhe Institute of Technology (KIT), Germany\\
mahdi.jafari2@kit.edu, rahul.sharma@kit.edu, sami.naim@student.kit.edu, christopher.gerking@kit.edu, reussner@kit.edu}
}



\maketitle

\begin{abstract}
Software architecture models capture early design decisions that strongly influence system quality attributes, including security. However, architecture-level security assessment and feedback are often absent in practice, allowing security weaknesses to propagate into later phases of the software development lifecycle and, in some cases, to remain undiscovered, ultimately leading to vulnerable systems. In this paper, we bridge this gap by proposing the generation of Abstract Penetration Test Cases (APTCs) from software architecture models as an input to support architecture-level security assessment. We first introduce a metamodel that defines the APTC concept, and then investigate the use of large language models with different prompting strategies to generate meaningful APTCs from architecture models. To design the APTC metamodel, we analyze relevant standards and state of the art using two criteria: (i) derivability from software architecture, and (ii) usability for both architecture security assessment and subsequent penetration testing. Building on this metamodel, we then proceed to generate APTCs from software architecture models. Our evaluation shows promising results, achieving up to 93\% usefulness and 86\% correctness, indicating that the generated APTCs can substantially support both architects (by highlighting security-critical design decisions) and penetration testers (by providing actionable testing guidance).
\end{abstract}

\begin{IEEEkeywords}
Software Architecture, Penetration Test Case Generation, Security by Design, Large Language Models
\end{IEEEkeywords}

\section{Introduction}
\label{sec:introduction}

Software architecture shapes the very foundation of each software system. As defined by Bass et al \cite{architecture_book}, a software architecture, like any design, can be viewed as a sequence of decisions made by architects during the early phases of the software development life cycle. Design decisions vary widely, ranging from structural decisions (e.g., defining components and interfaces that realize system functionality), to arrangement decisions (e.g., selecting architectural styles and patterns), and behavioral decisions (e.g., specifying connectors, communication mechanisms, and permitted dependencies among software elements) \cite{keim_taxonomy_2023}.

The design decisions made by software architects largely shape the quality attributes of the delivered system, particularly its security, which is our primary concern. As argued in \cite{Gennari2012MeasuringAS}, different classes of design decisions can significantly influence a system’s attack surface. For example, in the context of authorization, structural decisions such as where privilege and entitlement data is stored, how sessions and roles are managed, and when and where access checks are enforced can result in substantially different attack surfaces and security outcomes. This highlights the need to assess security weaknesses at the architectural level in a timely and cost-effective manner to verify that architectural decisions lead to an architecture that meets the required security level. Although architects typically make design decisions based on preliminary artifacts such as requirements, assumptions, and constraints, and in consultation with team members, they are rarely supported by systematic, security-risk–oriented feedback after selecting a concrete design alternative. As a result, they may lose the opportunity to assess whether the finalized architecture is sufficiently secure or requires revision. Consequently, potential weaknesses may remain hidden or be deferred to later development phases and may be uncovered only later through security assurance activities, such as penetration testing \cite{sarvejahani2025towards}.

To uncover weaknesses that are often overlooked in software systems, security testing is essential. Security testers must thoroughly investigate security risks to understand how the system responds to attacks. A widely used method for assessing software security is penetration testing. This type of security testing aims to simulate realistic attacks on the system in order to identify potential vulnerabilities \cite{arkin_software_2005}. Penetration testing is typically conducted by dedicated testing teams once the software has been fully developed and deployed. Studies have identified several significant drawbacks of traditional penetration testing, including its heavy reliance on domain knowledge and the expertise of human testers, who need to possess a deep understanding of the system. Additionally, issues that are discovered late in the testing process can lead to costly and time-consuming mitigation \cite{sarvejahani2025towards}. Consistently, another exploratory study involving 26 industry professionals found that many practitioners view penetration testing as cumbersome and resource-intensive, leading them to forgo it \cite{explorary_on_sec_features}. These insights suggest the need for more systematic, developer-friendly methods for testing security features.

As recommended in \cite{arkin_software_2005} and \cite{sarvejahani2025towards}, one way to address the limitations of classical penetration testing is to distribute penetration-testing activities across the software development lifecycle, especially during the design phase. This ambitious goal enables early assessment of security weaknesses at the architectural level, helping architects understand the security implications of their design decisions in a timely manner and define mitigations before implementation begins. To support design-level security assessment by shifting penetration testing earlier, we first need a clear notion of what a penetration test case (PTC) means in this context. Based on our review of the state of the art, to the best of our knowledge, there is no structured definition for an architectural-level PTC. Therefore, as our first contribution, we propose a metamodel that defines an abstract penetration test case (APTC), where \emph{abstract} means the test case is specified at the architecture level as a high-level, technology-agnostic test hypothesis (e.g., target components, attack vector, and assessed threat/weakness), without committing to concrete exploits, tools, or implementation details. Our aim is to assess security by applying a set of pre-generated APTCs to a given architecture. As our second contribution, we conduct an empirical study to investigate how effectively large language models (LLMs) can generate meaningful APTCs from software architecture as the primary input artifact. To do so, we apply multiple prompting strategies, including zero-shot, one-shot, few-shot, and chain-of-thought prompting, and evaluate two widely used LLMs, GPT and Gemini models. Our contributions not only define APTCs in a reusable and well-structured manner but also demonstrate the potential of LLMs for architecture-based security understanding. Furthermore, the results indicate which additional architectural annotations may be required to enable more precise generation of APTCs in future work. 

In summary, this paper investigates how penetration testing concepts can be shifted to the design phase by proposing an APTC definition as a core artifact and outlining how LLMs can be incorporated to support the generation pipeline. We further evaluate how well LLMs assist this generation task through an expert-based assessment. To this end, we pose the following research questions:
\begin{enumerate}[label=\textbf{RQ\arabic*:}, leftmargin=*]
  \item \label{rq:one} How should an abstract penetration test case (APTC) be defined to support architecture-level security assessment?
  \item \label{rq:two} To what extent can LLMs analyze and understand the security implications of a software architecture?
  \item \label{rq:three} To what extent can LLMs generate meaningful APTCs from software architecture models?
\end{enumerate}

This paper contributes \textbf{(C1)} a structured metamodel to define APTCs in a way to support architecture-level security assessment, \textbf{(C2)} an empirical evaluation of how effectively LLMs can generate meaningful APTCs from software-architecture artifacts using conducting multiple prompting strategies, from zero-shot to few-shot techniques, and \textbf{(C3)} an empirical evaluation of LLM's answers to point out limitations in LLM-generated outputs. This will also help determine which additional architectural annotations may be required at the metamodel level to enable more precise and actionable APTC generation.

The remainder of the paper is structured as follows. ~\Cref{sec:background} introduces the necessary background , including PCM and key concepts in penetration testing, LLMs, and prompt engineering. ~\Cref{sec:concept} presents our approach, including the APTC metamodel (to answer RQ1) and the LLM-driven APTC generation pipeline (to answer RQ2 and RQ3). ~\Cref{sec:evaluation} describes the evaluation setup, case studies, and results, along with threats to validity. Finally, ~\Cref{sec:conclusion} concludes the paper and outlines directions for future work.

\section{Background}  
\label{sec:background}

 \subsection{Palladio Component Model}
\label{sec:palladio}
The Palladio Component Model (PCM) \cite{BECKER20093} was proposed as part of the Palladio approach \cite{reussner2016modeling}, a software architecture simulator developed to analyse software at the architectural model stage for predicting Quality of Service (QoS) properties of component-based software architectures.

From the perspective of a software architecture team, PCM is considered an Architecture Description Language (ADL) used to structure and design component-based system architectures. To enable this, PCM provides five core models: the \textit{Repository model}, \textit{System model}, \textit{Resource Environment model}, \textit{Allocation model}, and \textit{Usage model}. Each of these models addresses a specific aspect that must be maintained in any architecture and enables QoS prediction. The \textit{repository model} outlines types of reusable components along with their interfaces and data structures. The \textit{system model}, also referred to as the \textit{assembly model}, wires component and interface instances to produce the system architecture. The \textit{resource environment model} defines resource containers, network connections between those defined containers, and processing resources. The \textit{allocation model} determines how assembly contexts (essentially component instances) are mapped to resource containers. Lastly, the \textit{usage model} focuses on user interactions by analyzing their workloads, scenarios, and behavior within the system. 

\subsection{Penetration Testing}
\label{sec:pentesting}

Ensuring the security of a software system is not only about adding defenses, countermeasures, and best practices; it also requires validating them through testing. A key method used for this assessment is called penetration testing. Penetration testing, also called pentesting, involves simulating real-world attacks to evaluate the risks associated with potential security breaches. During a pentest, testers not only identify vulnerabilities that could be exploited by attackers but also attempt to exploit these vulnerabilities, when possible, to determine what information or assets might be compromised following a successful attack \cite{weidman2014penetration}. According to the Penetration Testing Execution Standard (PTES)\cite{ptes_standard}, penetration testing is a multi-stage process consisting of four main phases: intelligence gathering, threat modeling, vulnerability analysis, and exploitation. These interconnected phases enable testers to understand the system being tested, identify vulnerabilities, and gain access to them. After completing a specific test, the tester documents all findings in a report known as the test report.




\subsection{Large Language Models and Prompt Engineering}
\label{sec:llms}
LLMs are deep neural networks, typically based on the Transformer architecture~\cite{vaswani2017attention}, trained on large-scale text corpora using self-supervised learning objectives such as next-token prediction. Through this training process, LLMs learn statistical patterns of language, enabling them to perform a wide range of tasks including text generation, summarization, reasoning, and code synthesis~\cite{wei2022emergent}. Recent models, such as \textit{GPT-5.2} and \textit{Gemini-3-Pro}, demonstrate strong generalization capabilities and can perform new tasks without task-specific fine-tuning. Instead, their behavior is guided through carefully designed input prompts. This paradigm makes LLMs particularly attractive for software engineering tasks, where structured reasoning over textual artifacts, such as architectural descriptions is required.

Prompt engineering~\cite{schulhoff2024prompt} refers to designing structured input instructions that guide an LLM to produce desired outputs. A prompt typically includes a task description, contextual information (e.g., a software architecture model), and optionally example input--output pairs that illustrate the expected output format. Prompting strategies differ in how much guidance they provide, particularly in whether demonstrations are included. In this work, we focus on three such strategies: \textit{zero-shot prompting}, which provides only the task description and context without examples; \textit{one-shot prompting}, which includes a single example input--output pair before the target task; and \textit{few-shot prompting}, which provides multiple examples to help the model infer patterns and produce more consistent, structured outputs~\cite{brown2020language}. Other prompting techniques also exist, such as \textit{chain-of-thought prompting}, which explicitly encourages step-by-step reasoning before producing a final answer~\cite{wei2022cot}. 





\section{Approach}
\label{sec:concept}
In this section, we present our approach for leveraging LLMs to generate APTCs and to evaluate their capabilities. We begin by introducing our running example, \textit{Maintenance}, a component-based scenario used in several prior architecture-based security analysis studies. To model the software architecture, we use PCM as our ADL. We chose PCM due to its semantic richness and extensibility, which facilitate further architecture-level analyses and simulation compared to other existing ADLs. This choice, however, does not imply that other ADLs are useless for this study. Our approach is therefore instantiated on PCM in this work, but the underlying idea is not inherently limited to PCM and could be adapted to other architecture description approaches that provide comparable structural and semantic information.

\begin{figure}
  \centering
  \includegraphics[width=\linewidth]{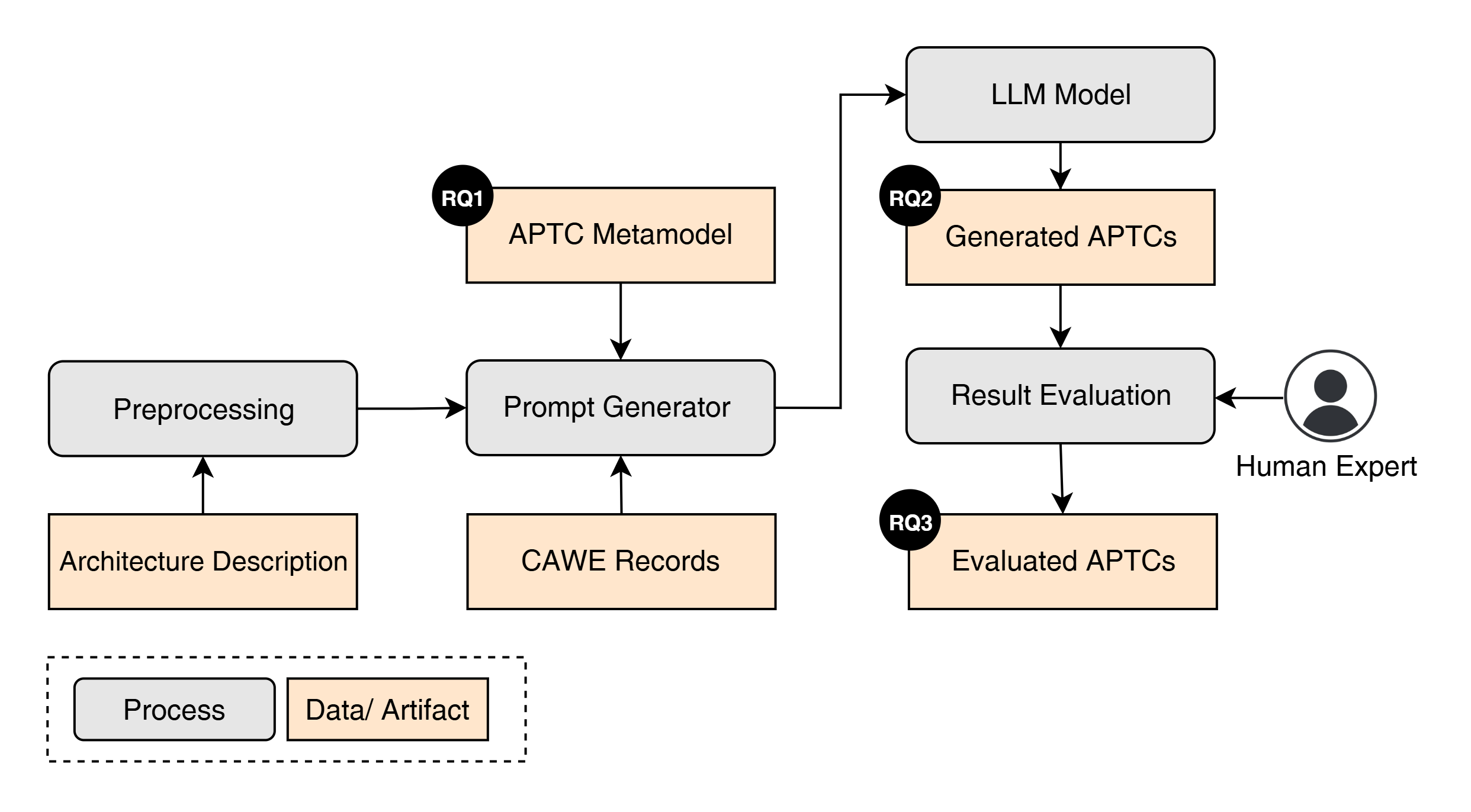}
  \caption{Overview of the proposed approach, from architecture description as input to APTC generation with LLMs and expert evaluation, mapped to the defined research questions.}
  \label{fig:pipeline}
\end{figure}

Next, we describe our first contribution: the metamodel that structures the definition of an APTC. With the running example described and the APTC definition in place, we move to the generation phase. At this stage, we pose a fundamental question that should be addressed first: \textit{What should an APTC at the architectural level aim to uncover?} To address this, we utilize the Common Architectural Weakness Enumeration (CAWE) \cite{cawe_cat} as the backbone. CAWE is a catalog that enumerates recurring security weaknesses rooted in software architecture, organized according to known security tactics and accompanied by mitigation guidance. To avoid confusion, note that Common Weakness Enumeration (CWE)\cite{noauthor_cwe_nodate} is not separate from CAWE. Rather, it is a curated subset of the well-known CWE that focuses specifically on architecture-rooted items. It currently lists 224 security-architecture weaknesses, which helps fill this gap. Accordingly, we propose generating APTCs that explicitly target weaknesses described in the CAWE catalog. Building on this idea, we leverage different prompting strategies, including zero-shot, one-shot, and few-shot, to develop prompts that elicit the LLM’s capabilities for APTC generation and to evaluate their effectiveness. We then proceed to the experiments, where we evaluate the approach using three experimental scenarios inspired by real-world systems using two widely used LLMs, OpenAI’s GPT and Gemini. \Cref{fig:pipeline} summarizes the approach and its mapping to the research questions.

\newcommand{\archelem}[1]{\emph{#1}}

\subsection{Running Example}
\label{sec:running}
We use the Maintenance scenario as a running example to present our study. The maintenance scenario was initially proposed by Al-Ali et al. \cite{modelingtrustcontractsindustry4.0} and developed in an Industry 4.0 environment. However, this does not restrict its application solely to that context; rather, it can be extended to any component-based systems \cite{archa_attack_propagation, attack_apth_ana}.

The scenario involves two organizations: a manufacturer (M) that owns and operates production machines, and a service contractor (S) responsible for maintaining and repairing those machines. During normal operation, an S technician can perform maintenance via a terminal, while the machine stores its data on an external storage system. However, the technician must not be allowed to access the machine’s log data under normal conditions, since the logs may contain sensitive information (e.g., employee names at M or detailed operational data). In the event of a machine failure, an S technician is permitted limited access to the log data for troubleshooting.

\begin{figure}[t]
  \centering
  \includegraphics[width=\linewidth]{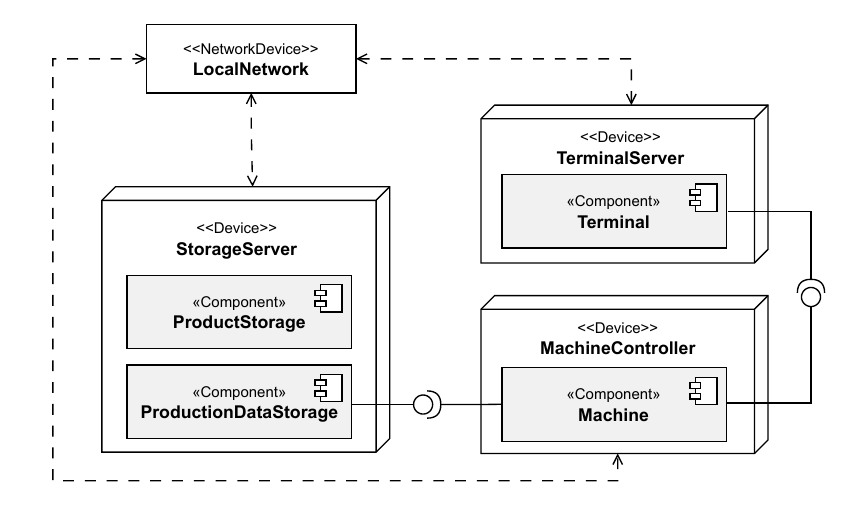}
  \caption{Simplified architecture of the running example, showing four components deployed across three devices and communicating via the LocalNetwork. The ProductStorage component contains confidential data and must be protected.}
  \label{fig:hms}
\end{figure}

 \Cref{fig:hms} illustrates the scenario’s components and their deployment. The system comprises four components: Terminal, Machine, ProductionDataStorage, and ProductStorage. These components are deployed across three hardware devices connected via a local network. ProductStorage, hosted on the storage device, contains confidential data about the production process and is therefore treated as the protected asset in this scenario. To simplify the access-control model, both the StorageServer and the TerminalServer are accessible only to users with the Admin role.

\subsection{Abstract Penetration Test Case Description}
\label{sec:ptcd}
To generate APTCs from the given software architecture, we first need to define what an APTC actually is. 
It is often beneficial to reference definitions from relevant resources or community standards. 
However, upon reviewing academic articles and pioneering standards in the domain, we found that, despite many impressive advantages of a structured definition, including benefits for automatic penetration testing, no formal or structured definition has been proposed for this purpose.
Moreover, in exploring the state of the art, we discovered that many common attributes and concepts are shared across various penetration standards and testing teams, even in the absence of a structured definition on the sheet. 
Therefore, given this missing part, we are motivated to fill this gap by proposing a metamodel to define an APTC to bridge APTC generation later on.

To structure our metamodel, we first analyzed state-of-the-art research papers relevant to our context. We then expanded the search to include well-established community standards. By consolidating these sources and focusing on the design phase, we refined the metamodel definition while acknowledging the limits of architecture-level information. Since our goal is to generate APTCs that can be used to identify weaknesses rooted in software architecture, we may omit test-case concepts described in prior work or standards that cannot be derived from architectural artifacts alone. Such concepts typically require evidence from later phases of the software life cycle (e.g., implementation or deployment), which are outside the scope of this paper. Therefore, we filtered candidate test-case concepts using two main criteria: \textit{(C1)} usefulness to both penetration testers and architects, and \textit{(C2)} realistic derivability from architectural artifacts.

After surveying the literature and shortlisting the most relevant work, we considered two studies\cite{xiong_model-driven_2010,durrwang_enhancement_2018} that are conducted to directly support PTC generation task. We then reviewed widely used standards and frameworks, namely the NIST Technical Guide \cite{scarfone_technical_2008}, the Open Source Security Testing Methodology Manual (OSSTMM3)\cite{herzog2006open}, the Penetration Testing Execution Standard (PTES)\cite{ptes_standard}, the ISSAF Standard\cite{oiss2006information}, and MITRE ATT\&CK\cite{al2024mitre} together with guidance from the Open Worldwide Application Security Project (OWASP) community. Collectively, these sources outline best practices for conducting penetration testing in broader software systems. This knowledge base forms the foundation for our first Contribution i.e., defining APTCs in this work.

Complementing these sources, we adopt the general notion of a \textit{test case} from ISO/IEC/IEEE 29119-1 \cite{noauthor_isoiecieee_nodate}, which standardizes core terminology for software testing. In this standard, a test case is a defined set of inputs, preconditions, and expected results, representing the lowest-level test input used to evaluate a test item. This general concept can also be applied equally to defining an APTC.

\begin{figure} [htb]
        \centering
        \includegraphics[width=.9\linewidth]{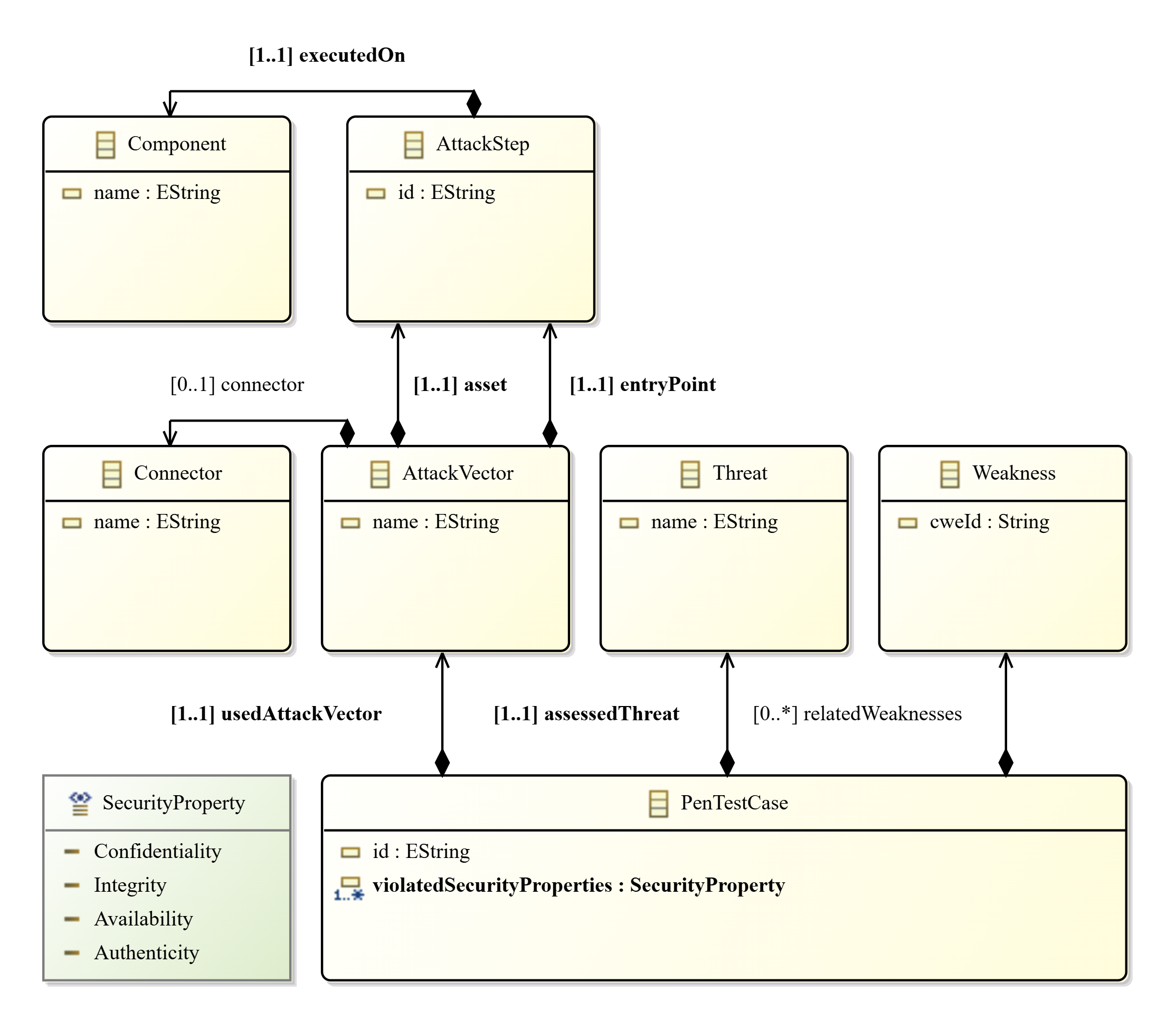}
        \caption{Abstract Penetration Test Case Metamodel.}
        \label{fig:ptcMetamodel}
    \end{figure}

\Cref{fig:ptcMetamodel} illustrates the metamodel we concluded to represent architecture-level APTCs.
The core concept is the \textit{PenTestCase}, which signifies an individual security test derived from the architecture. 
Each test case has an \textit{identifier} and is annotated with the security property violated by it, indicated by the attribute \textit{violatedSecurityProperties}. 
The enumeration \textit{SecurityProperty} captures the classical CIAA properties: \textit{Confidentiality}, \textit{Integrity}, \textit{Availability}, and \textit{Authenticity}. 
Each APTC is defined in relation to a specific \textit{threat} that is the focus of the test, \textit{weakness} that classifies the underlying weakness type targeted by the test, as well as an \textit{attack vector}. 
The association \textit{assessedThreat} links \textit{PenTestCase} to a \textit{Threat} element, representing a high-level adverse scenario (for example, ``Unauthorized disclosure of machine log data" or ``Manipulation of machine logs or machine state reports" in our running example), which is described by its name. The association \textit{relatedWeaknesses} links a test case to one or more \textit{Weakness} elements, indicating the CAWE weakness class(es) that the test case is intended to assess, i.e., the threat is assessed with respect to these weakness categories.
The association \textit{usedAttackVector} connects the test case to an \textit{AttackVector}, which denotes the specific means or path employed by a tester to execute the threat (such as “abuse of the technician terminal through credential theft or session hijacking” in our running example).
An attack vector is decomposed into two \textit{AttackSteps}. 
Each \textit{AttackStep} has an identifier and is referenced from the attack vector through two roles: the \textit{entryPoint} step indicates how the attacker initially gains a foothold, and the \textit{asset} step targets the final \textit{asset} whose compromise leads to the violation of the specified security property.
If the \textit{entryPoint} and the \textit{asset} are two different components, they are connected through a \textit{connector}. Although the current metamodel captures only an attack vector’s \textit{entryPoint} and \textit{asset} steps, it can be straightforwardly extended to represent more complex, multi-step attack vectors.

To specify where in the architecture the concrete attack occurs, each \textit{AttackStep} is linked to a \textit{Component} via the \textit{executedOn}. Each component is characterized by a unique name. Linking attack steps to components allows us to compute, for each APTC, the set and order of architectural elements involved in the attack (i.e., which components must be interacted with or compromised) and to validate whether the modeled attack is feasible given the connectivity between them.

In summary, the proposed metamodel captures the fact that each APTC assesses a specific threat by using a single attack vector. This attack vector is realized as a means or method for compromising the asset that starts at an entry-point step and ends at an asset step. Each step is bound to a concrete architectural component, and the overall effect is that certain security properties may be violated.
We emphasize that the proposed metamodel captures the fundamental elements that are broadly applicable across penetration test cases, rather than aiming to cover every possible variation found in the literature or in practice. This deliberate focus keeps the metamodel lightweight and aligned with our design-phase objective. At the same time, the metamodel is designed to be extensible so that additional concepts can be introduced as needed for specialized classes of test cases.


\subsection{Abstract Penetration Test Case Generation}
\label{sec:ptcg}

To enable security analysis from architectural models, we introduce a structured generation pipeline that derives APTCs from PCM specifications using LLMs. The pipeline preserves architectural semantics while enabling controlled, schema-compliant generation of security test artifacts.Given $(i)$ a PCM architectural model, $(ii)$ a selected LLM (e.g., \textit{GPT-5.2}), and (iii) a prompting configuration (zero-shot, one-shot, or few-shot), the generation process consists of three stages: $(1)$ security-oriented model serialization, $(2)$ constrained prompt construction, and $(3)$ schema-validated test case generation.

First, we serialize the PCM architecture with a focus on security. This means that the architecture is extracted and converted to a more suitable format for a security context. The output of the serialization contains sections, Overview of the architecture, interfaces and dependencies of components and resource environments of components. This serialization process reduces representational complexity by converting the structured PCM model into a linearized textual format suitable for LLM processing. At the same time, it ensures traceability by preserving exact architectural identifiers and refraining from introducing or modifying architectural elements.

The serialized architecture is embedded into a structured prompt template designed to enforce architectural fidelity and machine-processable output. The system role specifies, the exact JSON schema defining the structure of APTCs, mandatory reuse of architectural identifiers (e.g., ``Use the exact component and connector names from the provided architecture'', and explicit restrictions against introducing architectural elements not present in the serialized model. These constraints operationalize architectural consistency at the generation level and mitigate the risk of hallucinated components or invalid dependencies.

Depending on the prompting configuration, the input may additionally include one or more exemplar APTCs formatted according to the same JSON schema. In one-shot and few-shot settings, these examples provide structural guidance and reinforce schema adherence without altering the architectural context. The final LLM input therefore consists of $(i)$ the system instruction defining schema and constraints, $(ii)$ the serialized architecture representation, and $(iii)$ optional schema-compliant examples. \Cref{lst:prompt_simplified} shows a simplified base prompt. In the actual experiments, this prompt was adapted according to the prompting strategy used (e.g., zero-shot, one-shot, or few-shot).

\begin{lstlisting}[style=aptcjson, captionpos=b,
caption={Simplified base prompt used for APTC generation.},
label={lst:prompt_simplified}]
{
  "SystemRole": "You are a cybersecurity expert specializing in security-by-design. Your task is to analyze software architectures modeled in the Palladio Component Model (PCM) and generate Abstract Penetration Test Cases (APTCs).",
  "TaskDescription": "Given a PCM architecture description in Security Analysis View format, generate one architecture-level APTC for each of the following CWEs to assess whether the weakness exists in the architecture: CWE-284, CWE-285, CWE-862, CWE-863, and CWE-272.",
  "OutputFormat": "Return ONLY valid JSON (conforming to the schema below). Do not include markdown formatting, explanations, or conversational text.",
  "Constraints": [
    "Use exact component and connector names from the provided architecture description; do not invent names.",
    "Use only the CWEs listed above; do not introduce additional CWEs.",
    "If information is insufficient, set applicability to \"uncertain\" (or \"not_applicable\" per schema) and state the missing information in the appropriate field."
  ]
}
\end{lstlisting}

 Based on the constructed prompt, the LLM generates multiple APTCs in the predefined JSON format. Each generated test case is required to explicitly reference architectural entities defined in the PCM model, describe a security-relevant interaction grounded in the architectural structure, and conform syntactically to the predefined JSON schema. To enable automated integration into architecture-centric security workflows, all generated outputs are validated against the JSON schema. This validation step ensures structural correctness and tool compatibility for downstream analysis. By combining model-to-text serialization, constrained prompting, and schema validation, the pipeline maintains architectural traceability while leveraging the generative capabilities of LLMs for systematic derivation of architecture-based security test cases.

\section{Evaluation}
\label{sec:evaluation}

\subsection{Evaluation Design}
\label{sec:evaluationdesign}
We structure our evaluation according to the Goal-Question-Metric (GQM) approach~\cite{Basili1994TheGQ} and employ multiple experimental scenarios. Inspired by real-world applications, these scenarios serve both to demonstrate the applicability of the approach and to support the concrete measurement of the defined metrics within a well-defined evaluation framework. Following GQM, we define one evaluation goal, three evaluation questions, and two metrics to assess the proposed approach.

\begin{enumerate}[label=\textbf{G:}, leftmargin=*]
    \item  Evaluate whether LLMs can generate meaningful, architecture-grounded APTCs from PCM artifacts for a fixed set of CAWE entries, and identify the main limitations.
\end{enumerate}

To investigate this goal systematically, we define the following three evaluation questions:
\begin{enumerate}[label=\textbf{Q\arabic*:}, leftmargin=*]
    \item Do the generated APTCs conform to the required output format and correctly reference architectural elements?
    \item Do the generated APTCs correctly target the specified CAWE weaknesses at the architecture level?
    \item Are the generated APTCs useful to software architects and penetration testers?
\end{enumerate}

We answer these questions using two metrics. \textbf{\textit{M1 (Correctness)}} assesses whether an APTC (a) conforms to the required output format, the APTC metamodel, (b) is grounded in the given architecture by referencing only elements present in the input, and (c) matches the intended CAWE weakness. \textbf{\textit{M2 (Usefulness)}} assesses whether an APTC is actionable for software architects and penetration testers, i.e., whether it provides reasonable guidance to assess the targeted weakness in either software architecture (from an architectural perspective) or software system (from a pentester perspective). 

As a final step in the evaluation, we assessed the generated APTCs using two methods: (1) a purely expert-based evaluation and (2) an LLM-assisted expert-based evaluation. In the first, practitioners involved in this study were provided with the relevant artifacts, including the PCM models of the scenarios, the generated APTCs, and the metric definitions, and were asked to assign binary scores for correctness and usefulness. In the second, \textit{GPT-5.2} was given the same artifacts and instructed to assess the generated APTCs according to the same definitions. Its assessments were then reviewed manually by the practitioners. The unified results from both evaluation methods were considered as the final set and used to derive additional insights, such as success rates for comparing prompting strategies and LLMs.

The files of the evaluation and the implemented code used to generate the APTCs are available at the replication package~\cite{mahadi_2026}

%





\subsection{Experimental Scenarios}
\label{sec:casestudy}
To evaluate our approach, we use three real-world-inspired scenarios, including the running example and two additional scenarios. \textbf{\textit{(1) ABAC-Banking}} is based on the evaluation case study by Seifermann et al. \cite{SEIFERMANN2022111138}. It models a data-flow–based ABAC banking system with multiple branches and customer types (celebrity vs. regular), where access decisions depend on user role, customer type, and location (e.g., staff in Asia cannot access U.S. operations). \textbf{\textit{(2) PowerGrid}} consists of a corporate/business network (including services such as a domain controller and call-center applications) connected via a VPN to an ICS network. The ICS network hosts DMS server/client components used to operate grid breakers. \textbf{\textit{(3) Maintenance} }, which is our running example introduced earlier.
All three case studies modeled using PCM resulted in four models for each, including the Repository Model, System Model, Resource Environment Model, and Allocation Model, which serve as inputs to our APTC generation pipeline.

\begin{table*}[!htb]
\centering
\caption{Expert evaluation results for GPT-5.2 and Gemini-3-Pro across three case studies and prompting strategies. Total/15 aggregates over the three case studies (Maximum 1 APTC per each of 5 considered CAWEs); Success Rate is Total/15 as a percentage.}
\label{tab:expert_eval_gpt52_gemini3_total_rate}

\small
\setlength{\tabcolsep}{5pt}
\renewcommand{\arraystretch}{1.15}

\begin{tabularx}{\textwidth}{ll*{3}{>{\centering\arraybackslash}X} >{\centering\arraybackslash}X >{\centering\arraybackslash}X}
\toprule
\textbf{Model} & \textbf{Metric} &
\textbf{Maintenance} & \textbf{PowerGrid} & \textbf{Bank} &
\textbf{Total/15} & \textbf{Success Rate} \\
\midrule

\multicolumn{7}{l}{\textit{Zero-shot prompting}} \\
\midrule
\multirow{2}{*}{GPT-5.2}
 & Correctness & 2/5 & 3/5 & 4/5 & 9/15  & 60.0\% \\
 & Usefulness  & 5/5 & 4/5 & 4/5 & 13/15 & 86.7\% \\
\multirow{2}{*}{Gemini-3-Pro}
 & Correctness & 4/5 & 2/5 & 4/5 & 10/15 & 66.7\% \\
 & Usefulness  & 3/5 & 2/5 & 5/5 & 10/15 & 66.7\% \\
\midrule

\multicolumn{7}{l}{\textit{One-shot prompting}} \\
\midrule
\multirow{2}{*}{GPT-5.2}
 & Correctness & 4/5 & 3/5 & 4/5 & 11/15 & 73.3\% \\
 & Usefulness  & 4/5 & 3/5 & 4/5 & 11/15 & 73.3\% \\
\multirow{2}{*}{Gemini-3-Pro}
 & Correctness & 5/5 & 4/5 & 4/5 & 13/15 & 86.7\% \\
 & Usefulness  & 5/5 & 5/5 & 4/5 & 14/15 & 93.3\% \\
\midrule

\multicolumn{7}{l}{\textit{Few-shot prompting}} \\
\midrule
\multirow{2}{*}{GPT-5.2}
 & Correctness & 4/5 & 4/5 & 3/5 & 11/15 & 73.3\% \\
 & Usefulness  & 4/5 & 4/5 & 4/5 & 12/15 & 80.0\% \\
\multirow{2}{*}{Gemini-3-Pro}
 & Correctness & 2/5 & 2/5 & 2/5 & 6/15  & 40.0\% \\
 & Usefulness  & 2/5 & 5/5 & 3/5 & 10/15 & 66.7\% \\
\bottomrule
\end{tabularx}
\end{table*}

\subsection{Evaluation Results}
\label{sec:ttv}
For the evaluation, we considered only five CAWEs from the CAWE catalogue, focusing on weaknesses related to authorization and access-control enforcement. We chose this subset to keep the outcomes measurable for expert evaluation and to enable a clear comparison across LLMs and prompting strategies. The results of applying our approach to the three case studies introduced earlier are summarized in Table~\ref{tab:expert_eval_gpt52_gemini3_total_rate}. Overall, incorporating LLMs with prompting strategies yields promising results, in some cases achieving up to 93\% usefulness and 86\% correctness, suggesting that the generated APTCs are largely meaningful and architecture-relevant. The generated APTCs support software architects by highlighting security-critical parts of the design, sometimes surfacing important design decisions, and sometimes indicating areas that require extra attention. For penetration testers, the APTCs are also valuable: they function as concrete hypotheses and provide a clear direction for what to test and how. ~\Cref{ptc_sample} shows an example APTC produced by our approach for the running example, targeting CWE-863 (Incorrect Authorization), which describes situations where a system performs an authorization check but applies it incorrectly, for example by using the wrong conditions, roles, or context when deciding whether access should be granted.

\begin{lstlisting}[style=aptcjson, captionpos=b,
caption={Example APTC generated for the Maintenance case study .},
label={ptc_sample}]
{
  "CAWE": "CAWE-863",
  "violatedSecurityProperty": "Confidentiality",
  "Threat": "Authorization logic permits log access based on technician role alone, ignoring the required machine-failure state.",
  "AttackVector": {
    "Name": "Incorrect Authorization Logic Execution",
    "Connector": "MachineTerminal",
    "EntryPoint": "TerminalComponent",
    "Asset": "ProductionStorageComponent"
  }
}
\end{lstlisting}

 From a correctness perspective, the APTC is well-formed, maps to the intended CAWE, is architecture-grounded, and entirely conforms to the APTC metamodel. From a usefulness perspective, it pinpoints a key server-side design decision for software architects, log access must be authorized by both the technician role and the machine’s operational state (Failure vs. Normal), and it is immediately actionable for penetration testing: attempt log access while the machine is not in failure state, and try to bypass the check by omitting or faking any failure-state indicator.

\subsection{Discussion}
\label{sec:ttv}

The proposed approach offers several strengths for both software design and subsequent penetration testing activities. The generated APTCs provide a richer artifact by capturing threat-related information together with additional attributes such as the related CAWE entry, attack vector, and traceability to architectural elements. This can help software architects identify security-critical parts of the architecture and anticipate potential exploit scenarios, thereby supporting architecture-level security assessment. In turn, architects may be better able to reflect on the underlying design decisions and revise them toward a more secure architecture. The generated APTCs may also support penetration testers by reducing manual effort in early pentesting phases, especially during intelligence gathering and attack vector derivation.

While most generated APTCs were evaluated as correct and useful, clearly highlighting security-critical parts of the architecture, several were generated incorrectly and/or were not useful. In these cases, we observed a few recurring reasons. Sometimes, the LLMs failed to identify the appropriate violated security property for the targeted weakness. For example, one APTC targets CWE-862 (“Missing Authorization”) but frames the issue as Integrity/Tampering with “unchecked execution of sensitive operations” without specifying the sensitive operation or the missing authorization gate. As written, it does not match the scenario’s core risk (unauthorized log access outside failure mode) and is too abstract to validate against the architecture. Another common issue was inventing architectural elements or referring to the wrong components or connectors.

We identify several threats to validity. Construct validity may be affected by the security-oriented serialization of PCM models and by the predefined JSON schema, both of which may omit or constrain relevant architectural information. Internal validity is influenced by the non-determinism of LLM-based generation, including variations in prompting strategy, model parameters, and example design; moreover, schema validation ensures structural correctness, but not semantic soundness or practical exploitability. External validity is limited by the selected architectural models and evaluated LLMs, which may restrict generalizability. Finally, conclusion validity depends on the evaluation methodology, particularly where assessments rely on expert judgment and may therefore introduce subjectivity.
\section{Conclusion and Future Work}
\label{sec:conclusion}
To support software architects in making security-related design decisions, we highlighted the absence of security risk assessment in the early architecture design phase. To address this gap, we first proposed a metamodel to structure the definition of an APTC. The idea was to enable the generation of APTCs that support both architects, toward more secure and reliable software architectures, and penetration testers, by providing a set of pre-generated APTCs that guide them toward the right direction and security-critical parts to test. To this end, we proposed an LLM-based approach to investigate how well this technology can contribute in this context, and we developed different prompts to guide the LLMs. Overall, the results show that the generated APTCs are promising and useful for both architects and penetration testers, while also indicating the applicability of the proposed APTC metamodel.

As future work, we plan to systematically investigate the root causes of incorrect or unhelpful APTCs and extend the study to additional CAWEs to broaden coverage. In addition, we will explore the incorporation of further security artifacts, particularly threat and attacker models, to better ground the generation process and support LLMs in producing more precise and useful test cases. This is particularly relevant because threat modeling is a core step in penetration testing, helping translate general testing objectives into prioritized and realistic attack plans by identifying relevant assets, threat agents, and attacker capabilities. Lastly, the potential use of generated APTCs for more systematic and automated architecture-level security assessment remains an open direction for future investigation.
\section{Acknowledgment}
\label{sec:conclusion}

This work was funded by the Topic Engineering Secure Systems of the Helmholtz Association (HGF) and by the Deutsche Forschungsgemeinschaft (DFG, German Research Foundation) - SFB 1608 - 501798263. This work was supported by KASTEL Security Research Labs, Karlsruhe.

\bibliographystyle{ieeetran}
\bibliography{bibliography}

\end{document}